\documentclass[twocolumn, doublespace, 12pt fonts]{aastex6}
\usepackage{amsmath}
\usepackage{float}
\shorttitle{Powering GRB plateaus through LAEMWs}
\shortauthors{}

\begin{document}

\title{To power the X-ray plateaus of gamma-ray bursts through larger amplitude electromagnetic waves}

\author{Shuang Du$^{1,2}$}
\affil{
$^{1}${State Key Laboratory of Nuclear Physics and Technology, School of Physics, Peking University, Beijing 100871, China}\\
$^{2}${Kavli Institute for Astronomy and Astrophysics, Peking University, Beijing 100871, China}\\}

\email{dushuang@pku.edu.cn}
\begin{abstract}
The origin of gamma-ray burst (GRB) X-ray plateau, especially the internal plateau, is still unclear, but it could be related to GRB's central engine of magnetar.
It is generally believed that the spin-down power of the magnetar is injected into forward external shock, however we propose here that most of the power will be dissipated behind the GRB jet
through larger amplitude electromagnetic wave (LAEMW).
Based on this proposal, the relevant physical conditions and observational implications are analyzed and discussed, and various kinds of X-ray light curves could be reproduced.
Although it is still a matter of debate about the chromatic multi-band afterglow in the standard external afterglow fireball model, we can explain naturally this feature under this proposal,
i.e., the electrons generating the X-ray plateau and emitting the optical afterglow are accelerated by different mechanisms.
It is emphasized that both the GRB jet and the spin-down wind should have significant observational consequences in the magnetar scenario, and should be focused equally in GRB physics.

\end{abstract}


\section{Introduction}\label{sec1}
Neutron stars (NSs)/ magnetars are widely considered to interpret gamma-ray burst (GRB) observations, such as prompt emission
\citep{1992Natur.357..472U,1994MNRAS.267.1035U,1992ApJ...392L...9D,1998ApJ...508L.113K,1998PhRvL..81.4301D,2001A&A...369..694S,2007MNRAS.380.1541B},
extended emission \citep{2008MNRAS.385.1455M,2011MNRAS.413.2031M,2013MNRAS.431.1745G,2017MNRAS.470.4925G}, X-ray flares \citep{2006Sci...311.1127D}
and X-ray plateaus (as well as optical rebrightening, \citealt{1998A&A...333L..87D,2001ApJ...552L..35Z,2013MNRAS.430.1061R,THJZ2019}).
Alternatively, these features can also be interpreted if GRB central engines are black hole (BH)-disc systems
\citep{1992ApJ...395L..83N,1993ApJ...405..273W,1993Natur.361..236M,1999ApJ...518..356P,2005ApJ...630L.113K,
2007MNRAS.376L..48R,2011MNRAS.417.2161B,2017A&A...605A..60B}.
However, it is worth noting that some observational facts related to the X-ray plateaus are yet explained well under neither the NS scenario nor the BH scenario.

Generally, the plateau followed by a power-law decay with an index of $q<\sim -3$ is called `internal plateau'
(e.g., GRB 070110, \citealt{2007ApJ...665..599T}; GRB 090515, \citealt{2010MNRAS.409..531R}).
For clarity, in this paper, we call those plateaus with $q>-3$ ordinary plateaus (e.g., GRB 060729, \citealt{2007A&A...469..379E}).
A number of models are proposed to explain both types of X-ray plateaus.
Some interpretations do not depend on specific central engines but invoke:
a jet with an appropriate distribution of bulk Lorentz factor \citep{1998ApJ...496L...1R,2007ApJ...665L..93U},
a structured GRB jet \citep{2006ApJ...641L...5E,2006ApJ...640L.139T,2009ApJ...690L.118Y},
a jet with evolving microphysical parameters \citep{2006MNRAS.369.2059P,2006A&A...458....7I}, a jet with delayed deceleration \citep{2006MNRAS.366L..13G,2012ApJ...744...36S,2015ApJ...806..205D},
a jet that exchanges with circumburst medium \citep{2007ApJ...655..973K,2007ApJ...660.1319S}, and a jet viewed slightly off-axis \citep{2019arXiv190705899B}.
Those models may account for the ordinary plateaus but usually do not work for the internal plateaus due to the steep decay segments.

The leading model to explain both the ordinary plateaus and internal plateaus suggests a long-lasting central engine.
Without considering the interaction between the assumed long-lasting magnetized wind and GRB jet,
\cite{2017ApJ...835..206L} proposed that the ordinary X-ray plateaus can be produced through the interaction of the wind with the shocked circumburst medium.
But there must be an interaction between the wind and the GRB jet, since the assumed Lorentz factor of the wind, $\sim 10^{6}$, is much larger than that of the GRB jet.
And there should be a very weak interaction between the wind and the circumburst medium, since the circumburst medium is swept up by the GRB jet.
To account for the internal plateau, we remind that energy dissipation of the continuous outflow should trace energy release of the central engine.
Let's have a brief review.

Under the long-lasting BH-disc central engine scenario \citep{2008Sci...321..376K},
the different segments of outflow, as well as the light curve, are related to the accretion of different parts of the GRB progenitor star,
 i.e., the X-ray plateau is just corresponding to the accretion of a certain region of the progenitor star.
But, this model can only be applied to collapsar.
Even if one believes that this long-lasting accretion is also true for short GRBs, as shown by their erratic rapidly variable prompt-emission light curves,
the accretion, as well as the GRB jets, should be intermittent and inhomogeneous.
Then a rapidly variable ``late prompt'' emission produced by the intermittent and inhomogeneous long-lasting outflow should be observed but not
the smooth plateau (e.g., GRB 070110,  \citealt{2007ApJ...665..599T}).
It is not rigorous to believe that the energy release of the intermittent and inhomogeneous jet is smoothly (e.g., \citealt{2007ApJ...658L..75G,2017A&A...605A..60B}).


Under the magnetar scenario, on one hand, the ordinary plateau would be powered by the
spin-down wind of stable magnetar with a certain braking index. On the other hand, the spin-down-induced gravitational collapse of unstable magnetar could account
for the steep decay of internal plateau phenomenologically \citep{2013MNRAS.430.1061R}.
However, it is generally believed that the energy of spin-down wind will be directly injected into the forward external shock (e.g., \citealt{1998A&A...333L..87D,2001ApJ...552L..35Z}),
so that the decay follows the internal plateau should be a normal decay (decay index $\sim -1.2$, \citealt{2006ApJ...642..354Z}) instead of a steep decay after the magnetar collapsing.
Because, either before or after energy injection, the X-ray is emitted by the electrons accelerated by forward external shock.
Some authors suggest that there may be some unknown internal dissipation in the magnetized winds \citep{2006MNRAS.372L..19F,2010ApJ...715..477Y}.

Given the natural correspondence between the collapse of magnetar and the steep decay of internal plateau,
we prefer the magnetar scenario. In Section 2, we introduce the applicability of GRB NS/magnetar model,
and propose that the GRB jet and the spin-down wind might evolve separately. In Section 3, we discuss some implications of this proposal.
Section 4 is discussion and summary.
Throughout this paper, parameters without primes are in the lab frame,
and parameters with primes are in the rest frame co-moving with the spin-down wind.

\section{Why is NS/magnetar needed?}
GRBs can be classified into two categories based on duration $T_{\rm 90}$  \citep{1993ApJ...413L.101K}: short GRBs with $T_{\rm 90}<2\;\rm s$
and long GRBs with $T_{\rm 90}>2\;\rm s$. Observations confirm that short GRBs at least originate from double NS mergers (GW 170817/GRB170817A association, \citealt{2017ApJ...848L..13A}),
and long GRBs originate from massive star collapses (e.g., GRB/type-Ic supernova associations, \citealt{1998Natur.395..670G,2003Natur.423..847H,2003ApJ...591L..17S}).
The central remnants of these catastrophes are still uncertain.
As illustrated in the Introduction, the BH-disc system should not generate a smooth X-ray plateau due to the inhomogeneous accretion, hence we turn to the NS central engine.

For an NS-NS merger, the merger remnant may be an NS or a BH, which depends on the total mass of the binary and the equation of state of NSs.
A precise measurement shows the upper limit on the rest mass of NS should satisfy $M_{\rm ToV}>1.97\; M_{\rm \odot}$  \citep{2013Sci...340..448A}.
GW 170817/GRB 170817A indicates $M_{\rm TOV}$ may be $\sim 2.2\; M_{\rm \odot}$  (\citealt{2017ApJ...850L..19M}, alternatively, see \citealt{2019NatAs.tmp..439C})
or greater \citep{2018ApJ...861..114Y}.
In this sense,  the maximum mass that a rotating NS can support will be $\sim 2.6\;M_{\rm \odot}$ or more.
Therefore, if the total mass distribution of extragalactic double NS systems is similar to that of in the Milky Way, i.e., $M_{\rm tot}\in (2.5\;\rm M_{\rm \odot}, 2.9\;\rm M_{\rm \odot})$
(\citealt{2016ApJ...831..150L,2018ApJ...854L..22S}; see \citealt{2016ARA&A..54..401O} for review),
the remnant of NS-NS merger can be a supramassive/hypermassive NS (even a stable NS).
For massive star collapse, since an NS can be born in a supernova explosion, the centre of the long GRB associated with the supernova can also be an NS (stable, supramassive, hypermassive).
Therefore, NS/magnetar scenario is reasonable at least for part of GRBs.

To avoid the difficulty of magnetar scenario mentioned in the introduction
that the standard energy injection under magnetar scenario can only produce the ordinary plateau followed by a power-law decay with an index of $q\sim -1.2$,
we suggest an improved scenario that most of the spin-down power of magnetar is usually not injected directly into the forward external shock but continuously dissipated behind
the GRB jet independently (mainly through large amplitude electromagnetic waves, LAEMWs, see below).
Notably, once LAEMWs can keep the energy dissipation of the spin-down wind tracing the spin-down power (which is stronger than the external-shock-induced X-ray emission of the GRB jet),
the collapse of the magnetar will be naturally corresponding to the steep decay of the internal plateau.
In this case, the GRB jet and the spin-down wind evolve separately,
so that both of the GRB jet and the spin-down wind should have significant observational consequences.

It is worth noting that this proposal does not depend on the types of plateaus and can be applicable for both the ordinary and the internal plateaus.
Interestingly, if one believes the X-ray transient CDF-S XT2 \citep{2019Natur.568..198X} is powered by a magnetar,
our proposal can also be consistent with the implication
that the energy release of the spin-down wind has nothing to do with the GRB jet.
In the next, we discuss some implications of our proposal.
We will focus on the internal plateaus, since the discussion can be naturally generalized to the ordinary plateaus.

\section{Implications of the idea}\label{sec.2}

\subsection{LAEMWs}\label{sec3.1}
In order to explain the observed excess in infrared radiation from Crab Nebula,
Usov proposed that synchrotron radiation from electrons accelerated in the field of LAEMWs (\citealt{1975Ap&SS..32..375U}, also see \citealt{1994MNRAS.267.1035U,1996MNRAS.279.1168M})
can explain this observation. In this section, we apply this concept to GRB X-ray internal plateaus.
Let's introduce LAEMWs first.

For a rotating NS with inclined dipole magnetic field in vacuum, the magnetic dipole radiation generated by this magnetic dipole can be solved analytically under a given accuracy.
But in reality, the NS is contained in a magnetosphere filled with plasma due to unipolar induction effect \citep{1969ApJ...156...59G}.
Therefore, this low-frequency magnetic dipole radiation will be screened at near distance.

Nevertheless, the rotational energy of the NS must be extracted by the approximatively isotropic outflow (spin-down wind) consisting of plasma and magnetic field frozen inside.
Certainly, since there is an inclination angle between the rotation axis and the magnetic axis of the NS, the frozen magnetic field in the spin-down wind along
the rotation axis (as well as the GRB jet) can be alternating-direction (i.e., striped magnetic field, see, e.g., \citealt{1990ApJ...349..538C,1999ASPC..190..153U,2001A&A...369..694S}).
As the spin-down wind moves away from the NS and the wind density decreases,
there is a moment that displacement currents can not be screened any more and an induced electric field occurs.
Eventually, the striped magnetic field is transformed into low-frequency electromagnetic waves, i.e., LAEMWs, in far field.

\subsection {How does the spin-down wind dissipate?}
In this subsection, we discuss how can the spin-down wind be dissipated through LAEMWs.
To generate an almost flat X-ray plateau and a steep decay, energy of the spin-down wind should be released rapidly and smoothly.
Therefore, (i) the spin-down wind should be highly magnetized initially, i.e.,
\begin{eqnarray}\label{eqs}
\sigma=\frac{B^{2}}{4\pi \Gamma_{\rm w}\rho c^{2}}\gg 1,
\end{eqnarray}
\begin{eqnarray}
\frac{B^{2}}{4\pi}vS\approx L_{\rm sd},
\end{eqnarray}
where $\sigma$ is the magnetization, $B$ is the magnetic field strength, $\rho$ is the mass density, $v$ is the speed of the spin-down wind, $S$ is the cross area,
$c$ is the speed of light, $L_{\rm sd}$ is the spin-down power, and $\Gamma_{\rm w}$ is the bulk Lorentz factor of the spin-down wind;
(ii) the spin-down power should be balanced by magnetic-energy dissipation rate, $L_{\rm dis}$, approximatively, i.e.,
\begin{eqnarray}\label{c1}
\frac{L_{\rm sd}}{L_{\rm dis}}=\frac{B^{2}Sv}{4\pi n_{\pm }P_{\rm e}Sl}\sim 1,
\end{eqnarray}
where $n_{\pm}$ is the number density of accelerated electrons (and positrons), $P_{\rm e}$ is the synchrotron radiation power of a single electron
and $l$ is the distance that the wind travels per second \footnote{Strictly, $l$ should be the radiative damping length of LAEMWs.
However, to result in a plateau instead of a rising light curve during the hydrodynamic evolution time scale of the spin-down wind, $\sim r_{\rm d}/\Gamma_{\rm w}^{2}c$,
with $r_{\rm d}$ the radius of the dissipation region and $\Gamma_{\rm w,0}$ the Lorentz factor of the spin-down wind,
we need $l$ to be small enough, i.e., $l/c\leq r_{\rm d}/\Gamma_{\rm w,0}^{2}c$, so that the energy of the LAEMWs would be dissipated near $r_{\rm d}$.
Note that $r_{\rm d}\sim 10^{12}\;\rm cm$ (see later in this sub-section) and the order of magnitude of $\Gamma_{\rm w,0}$ should be $\sim 10$ (see section \ref{sect3.4}),
there is $l/c\sim 1$. This argument also is consistent with the estimations of \cite{1978A&A....65..401A} and \cite{1994MNRAS.267.1035U}.
Therefore, we simply consider $l$ as the distance that the wind travels per second.};
(iii) when the magnetic energy in the volume $Sv$ is dissipated totally, electrons should also be cooled down,
 i.e., the radiative lifetime of these electrons, $\tau$, satisfies
\begin{eqnarray}
\tau=\frac{ E_{\rm e}}{P_{\rm e}}< \sim 1\;\rm s,
 \end{eqnarray}
 where $E_{\rm e}$ is the energy of a single electron.
Condition (i) enables the energy of the spin-down wind almost to be released totally once the magnetic energy is completely dissipated.
Conditions (ii) and (iii) guarantee that the magnetic energy can be dissipated rapidly and the energy of the accelerated electrons does not accumulate,
so that the corresponding light curve keeps flat.

For magnetic-field-dominated relativistic wind, according to conditions (ii) and (iii), we have
\begin{eqnarray}\label{eqa}
\frac{B_{\rm d}^{\;2}}{4\pi n_{\pm}P_{\rm e}}=\frac{B_{\rm d}^{'2}}{4\pi n_{\pm}^{'}P_{\rm e}^{'}\Gamma_{\rm w}} \sim 1\;\rm s,
\end{eqnarray}
and
\begin{eqnarray} \label{eqb}
 \frac{E_{\rm e}^{'}}{\Gamma_{\rm w} P_{\rm e}^{'}}\approx\frac{5\times 10^{8}}{\Gamma_{\rm w} \gamma_{\rm e}B_{\rm d}^{'2}}<1\;\rm s,
\end{eqnarray}
where $\gamma_{\rm e}$ is the Lorentz factor of electrons.

Note that the plateau usually appears at $t_{\rm a}\sim 100\;\rm s$ after the burst,
the spin-down wind at least dissipates at the place $r_{\rm d}=ct_{\rm a}\sim 10^{12}\;\rm cm$ away from the source.
Since magnetic field in the spin-down wind is dominated by the toroidal component at a large distance from the source
and the dipole magnetic field in the light cylinder, the magnetic field strength at $r_{\rm d}$ is (see also \citealt{1994MNRAS.267.1035U})
\begin{eqnarray}\label{eq2}
B_{\rm d}&\approx& B_{\rm dip}\left ( \frac{R_{\ast}}{r_{\rm lc}} \right )^{3}\left ( \frac{r_{\rm lc}}{r_{\rm d}} \right )\nonumber\\
&=&1.3\times 10^{7}\left ( \frac{r_{\rm d}}{10^{12}\;\rm cm} \right )^{-1}\nonumber\\
&&\times\left ( \frac{B_{\rm dip}}{3\times 10^{14}\;\rm Gs} \right )\left ( \frac{P}{1\;\rm ms} \right )^{-2}\;\rm Gs,
\end{eqnarray}
where $B_{\rm dip}$, $r_{\rm lc}=cP/2\pi$ and $P$ are the surface magnetic field strength, the light cylinder radius and spin period of the magnetar, respectively.
The method to derive the typical values of $B_{\rm dip}$  and $P$ can be found in \cite{Du2016,Du2019}.
For simplicity, radius of the magnetar that $R_{\ast}=10^{6}\rm \; cm$ is adopted.
To get equation (\ref{eq2}), we also assume that the magnetic energy release is not important before the spin-down
wind reaching $r_{\rm d}$.
The reason is that instabilities which cause magnetic field dissipation need time to propagate.
The typical propagating time scale is the Alfv\'{e}n crossing time over the length scale that magnetic field in the outflow changes.
This gives a distance from the source $ r\sim 10^{12}\;\rm cm$ \citep{2001A&A...369..694S} beyond which the magnetic energy release becomes important.
It is clear that $r\sim 10^{12}\;\rm cm$ matches $r_{\rm d}$ well.

The typical energy of synchrotron emission is
\begin{eqnarray}\label{eq1}
\varepsilon=1.7\left ( \frac{\gamma_{\rm e}^{'}}{10^{2}} \right )^{2} \left ( \frac{B_{\rm d}}{10^{7}\;\rm Gs} \right )\;\rm keV.
\end{eqnarray}
Combining equations (\ref{eq2}) and (\ref{eq1}), we find that when $B_{\rm d}\sim 10^{7}\;\rm Gs$ and $\gamma_{\rm e}^{'}\sim 10^{2}$,
electrons can emit X-ray photons at $r_{\rm d}\sim 10^{12}\; \rm cm$. Meanwhile equation (\ref{eqb}) can be easily satisfied.
According to the fast dissipation condition, i.e., equation (\ref{eqa}), if most electrons in the spin-down wind are accelerated, there is
\begin{eqnarray}\label{eq3}
n_{\pm}=7.2\times 10^{9}\left ( \frac{\gamma_{\rm e}^{'}}{10^{2}} \right )^{-2}.
\end{eqnarray}

Now, the question is what mechanism accelerates these electrons.
We have argued that the spin-down wind is continuously dissipated behind the GRB jet,
so that without internal collisions \citep{2011ApJ...726...90Z}
the ordered striped magnetic field in the spin-down wind may be hard to be dissipated into X-ray emission violently (see \citealt{2016JPlas.82..0502P} for review)\footnote{
To convert the magnetic energy of spin-down wind into kinetic energy to produce a GRB,
some other mechanisms are proposed  (e.g., \citealt{2002A&A...391.1141D,2010ApJ...725L.234L}).
But, the prediction of these mechanisms is not supported by the observation that GRBs are produced by
relativistic jets rather than approximatively isotropic spin-down winds \citep{2018Natur.561..355M,2019Natur.565..324I}.
Therefore, we do not consider these mechanisms.}.
As introduced in subsection \ref{sec3.1}, LAEMWs may be another way.
The critical condition for the emergence of LAEMWs is that the number density, $n_{\pm}$, decreases to the Goldreich-Julian density \citep{1994MNRAS.267.1035U}
\begin{eqnarray}\label{eq6}
n_{\rm cr}&=&1.0\times 10^{9} \left ( \frac{r_{\rm d}}{10^{12}\;\rm cm} \right )^{-1}\nonumber\\
&&\times\left ( \frac{B_{\rm dip}}{3\times 10^{14}\;\rm Gs} \right )\left ( \frac{P}{1\;\rm ms} \right )^{-3}\;\rm cm^{-3}.
\end{eqnarray}
Since $n_{\pm}\propto r^{-2}$ and $n_{\rm cr}\propto r^{-1}$,
comparing equations (\ref{eq3}) and (\ref{eq6}), one can find the dissipation of magnetic energy via
LAEMWs will be important when the spin-down wind reaches $\sim r_{\rm d}$.
It is worth mentioning that before the first data point is observed, Chandra Telescope had pointed at
CDF-S XT2. That's to say, the X-ray emission of CDF-S XT2 appears suddenly without a gradual brightening as shown in its light curve.
This feature is consistent with our scenario that the spin-down wind dissipates immediately when it reaches $r_{\rm d}$.

\subsection{Gamma-ray photons and X-ray photons, which is first?}
In the above, we have shown that the spin-down wind is dissipated at least at $r_{\rm d}\sim 10^{12}\;\rm cm$, since X-ray plateau usually appears at $\sim 100\;\rm s$ after the burst.
However, the time delay, $\Delta t\approx1.7\;\rm s$, between GW 170817 and GRB 170817A indicates that the prompt emission occurs at \citep{1994ApJ...430L..93R,2017ApJ...848L..34M}
\begin{eqnarray}
R_{\gamma}=2c\Delta t\Gamma_{\rm jet}^{2}=1.0\times 10^{15} \left ( \frac{\Gamma_{\rm jet}}{10^{2}} \right )^{2}\;\rm cm,
\end{eqnarray}
if the Lorentz factor of the relativistic jet launched from the binary NS merger satisfies $\Gamma_{\rm jet}\gg 1$ initially.
Accordingly, if the dissipation of the spin-down wind follows the generation of gamma-ray photons, there is $r_{\rm d}\sim R_{\gamma}$.
Therefore, the parameters derived through equations (6), (7) and (8) turn to
\begin{eqnarray}\label{eq9}
B_{\rm d}&\approx& B_{\rm dip}\left ( \frac{R_{\ast}}{r_{\rm lc}} \right )^{3}\left ( \frac{r_{\rm lc}}{r_{\rm d}} \right )\nonumber\\
&=&1.3\times 10^{4}\left ( \frac{r_{\rm d}}{10^{15}\;\rm cm} \right )^{-1}\nonumber\\
&&\times\left ( \frac{B_{\rm dip}}{3\times 10^{14}\;\rm Gs} \right )\left ( \frac{P}{1\;\rm ms} \right )^{-2}\;\rm Gs,
\end{eqnarray}
\begin{eqnarray}\label{eq7}
\varepsilon=1.7\left ( \frac{\gamma_{\rm e}^{'}}{3.3\times 10^{3}} \right )^{2} \left (\frac{B_{\rm d}}{10^{4}\;\rm Gs} \right )\;\rm keV,
\end{eqnarray}
\begin{eqnarray}\label{eq8}
n_{\pm}=7.2\times 10^{6}\left ( \frac{\gamma_{\rm e}^{'}}{3.3\times 10^{3}} \right )^{-2}.
\end{eqnarray}
However, the condition of emerging LAEMWs is still satisfied.
Because, at this time, the critical number density is
\begin{eqnarray}\label{eq10}
n_{\rm cr}&=&1.0\times 10^{6} \left ( \frac{r_{\rm d}}{10^{15}\;\rm cm} \right )^{-1}\nonumber\\
&&\times\left ( \frac{B_{\rm dip}}{3\times 10^{14}\;\rm Gs} \right )\left ( \frac{P}{1\;\rm ms} \right )^{-3}\;\rm cm^{-3}.
\end{eqnarray}
One still has $n_{\rm cr}\sim n_{\pm}$.

It is worth noting that
if the $\sim 1.7\;\rm s$ time delay between GW 170817 and GRB 170817A is caused by delayed jet launch
or acceleration of jet from non-relativistic case to extreme relativity \citep{1993Natur.361..236M}, there is no need to demand $R_{\gamma}\sim 10^{15}\;\rm cm$ .
But one can't rule out the possibility that the initial X-ray photons due to the dissipation of spin-down wind can be detected earlier than the GRB prompt emission
 (e.g., $r_{\rm d}\sim 10^{12}\;\rm cm$, $R_{\gamma}\sim \;\rm 10^{15}\;\rm cm$), as long as the initial interval, $\Delta L$, between the spin-down wind and the jet satisfies
 \begin{eqnarray}
 \Delta L&<&\frac{R_{\gamma}}{2\Gamma_{\rm jet}^{2}}\nonumber\\
 &=&5.0\times 10^{10}\left ( \frac{R_{\gamma}}{10^{15}\;\rm cm} \right )\left ( \frac{\Gamma_{\rm jet}}{10^{2}} \right )^{-2}\;\rm cm.
\end{eqnarray}
Therefore, this feature provides a chance to test our proposal.

\subsection{Interaction between the spin-down wind and GRB jet}\label{sect3.4}

When the spin-down wind reaches $r_{\rm d}$, the magnetic energy would not be totally dissipated to power the X-ray plateau.
Part of the magnetic energy may convert to the kinetic energy of the remnant spin-down wind. 
Energy conservation gives
\begin{eqnarray}\label{eq17}
\eta(1+\sigma_{0})\Gamma_{\rm w}= (1+\sigma_{\rm d})\Gamma_{\rm w,d},
\end{eqnarray}
where $\sigma_{0}$ and $\sigma_{\rm d}$ are the magnetization before and after the spin-down wind dissipation,
$\Gamma_{\rm w}$ and $\Gamma_{\rm w,d}$ are the Lorentz factors of the spin-down wind before and after the dissipation,
and $\eta$ is the efficiency with which magnetic energy is converted into kinetic energy. Since the circumburst medium is swept up by the GRB jet,
we assume that $\Gamma_{\rm w}$ and $\Gamma_{\rm w,d}$ are constants.
From equation (\ref{eq17}), one has
\begin{eqnarray}\label{A18}
\Gamma_{\rm w,d}\sim 10^{2}\left ( \frac{\eta }{0.1} \right )\left ( \frac{\Gamma_{\rm w}}{10} \right )
\end{eqnarray}
with $\sigma_{0}\sim 10^{2}$ and $\sigma_{\rm d}\sim 1$.

\begin{figure}
 \centering
 \includegraphics[width=0.4\textwidth]{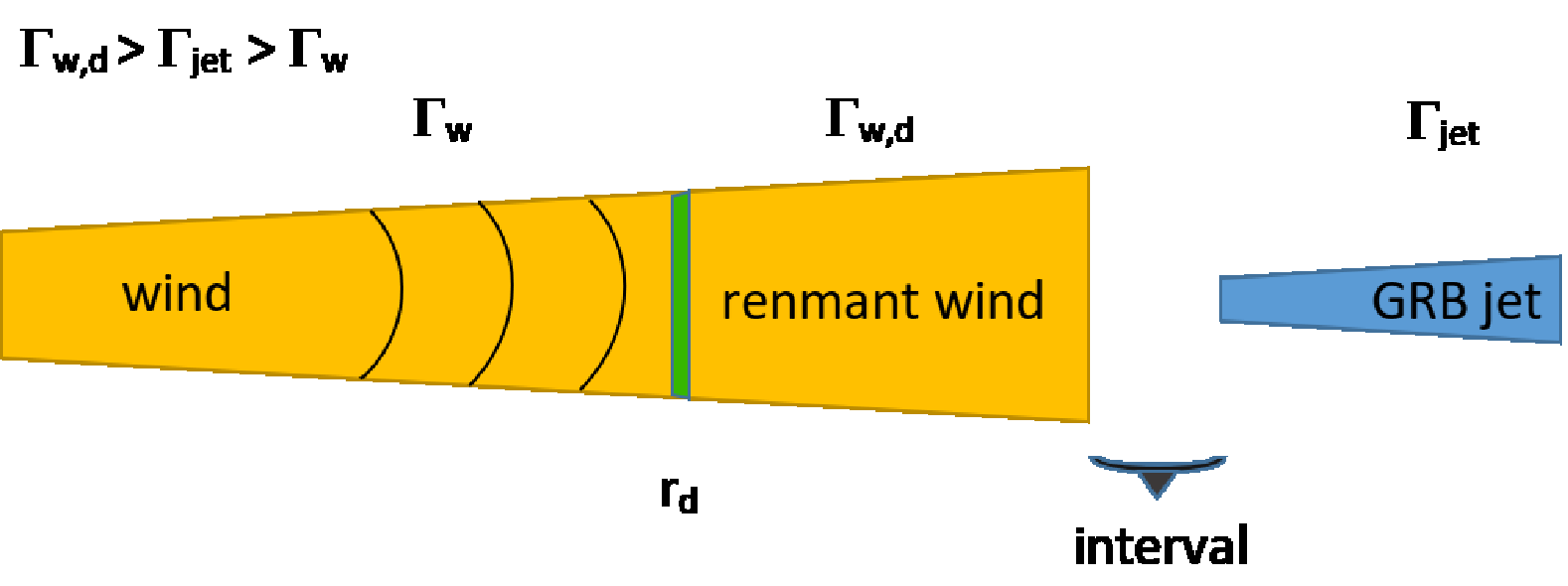}
 \includegraphics[width=0.4\textwidth]{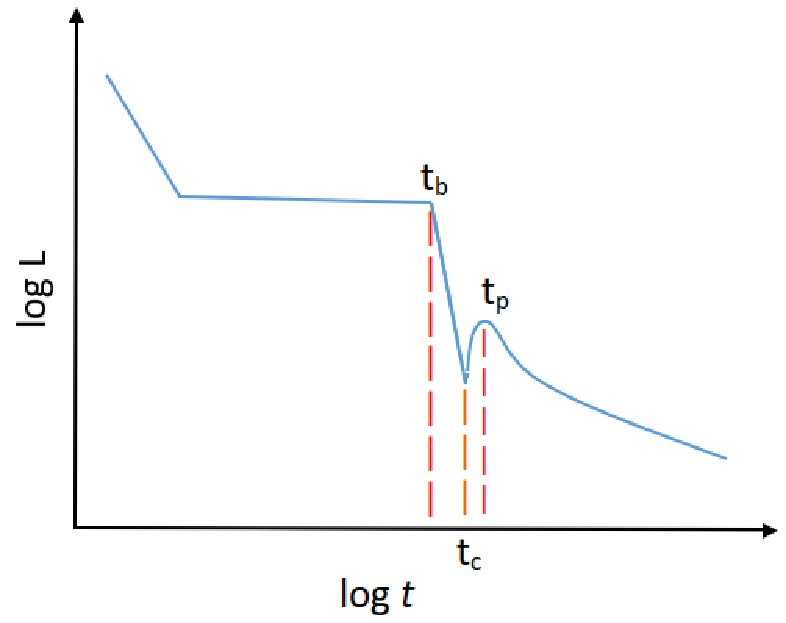}
 \caption{Schematic diagram of the relation between the spin-down wind and the GRB jet under $\Gamma_{\rm w,d}>\Gamma_{\rm jet}>\Gamma_{\rm w}$.
  The black curves denote the magnetic lines. Vertical bar at $r_{\rm d}$ is the place in which the spin-down wind dissipates.
  The approximative isotropic spin-down wind is cut into trapezoid to fit the size of the image (these stipulations are also applied for Figure \ref{geo} and Figure \ref{fig.3}).
Since there is $\Gamma_{\rm w}<\Gamma_{\rm jet}$ initially, an interval will exist between the spin-down wind and GRB jet.
  When the spin-down wind reaches $r_{\rm d}$, the magnetic-energy dissipation results in the X-ray plateau and leads to the acceleration of the remnant spin-down wind.
  Consequently, the Lorentz factor of the remnant spin-down wind may satisfy $\Gamma_{\rm w,d}>\Gamma_{\rm jet}$.
  After the last spin-down wind (far left of the bigger trapezoid) reaches $r_{\rm d}$  (at time $t_{\rm b}$), steep decay begins.
  And then, the faster remnant wind eventually catches up with the GRB jet at time $t_{\rm c}$, so that a flare-like X-ray bump appears.}
  \label{fig1}
\end{figure}

Note that there could be an interval between the spin-down wind and GRB jet, e.g., $\Gamma_{\rm w}<\Gamma_{\rm jet}$ due to the initial mass loading of the wind
or the delayed spin-down wind launch induced by fall-back accretion \footnote{The evolution of $\Gamma_{\rm jet}$ can be found in \cite{HGDL}.}. If the acceleration of the remnant spin-down wind satisfies $\Gamma_{\rm w,d}>\Gamma_{\rm jet}$ (see equation \ref{A18}),
the remnant spin-down wind will catch up with the GRB jet (i.e., the case shown in the upper panel of Figure \ref{fig1}),
so that a collision can happen.
Through the collision, the forward shock propagating in the GRB jet always can be produced, however the emergence of reverse shock depends on the
residual magnetic field in the spin-down wind.
The critical condition is that the pressure of the shocked jet matter equals to the
magnetic pressure of the spin-down wind \citep{2018pgrb.book.....Z}, i.e.,
\begin{eqnarray}\label{A1}
\frac{4}{3}\Gamma_{\rm w,jet}^{2}\Gamma_{\rm jet}n_{\rm w}m_{\rm p}c^{2}\sim \frac{4}{3}\Gamma_{\rm w,d}^{2}n_{\rm w}m_{\rm p}c^{2}\simeq \frac{B^{2}}{8\pi},
\end{eqnarray}
where $\Gamma_{\rm w,jet}$ is the relative Lorentz factor between the wind and jet.
For rough dimensional analysis,
we assume that the ratio of the baryon number density of the spin-down wind to the baryon number density of the jet, $n_{\rm w}/n_{\rm jet}$,
is approximate to the ratio of the spin-down luminosity to the jet power.
Therefore, combining equation (\ref{eqs}) and equation (\ref{A1}) gives
\begin{eqnarray}\label{eq18}
\sigma_{\rm cri}&\sim &\frac{8}{3}\Gamma_{\rm w,d}^{2}\frac{n_{\rm w}}{n_{\rm jet}}\nonumber\\
&\approx&7\times 10^{5} \left ( \frac{\Gamma_{\rm w,d}}{50} \right )^{2}\left ( \frac{n_{\rm jet}/n_{\rm w}}{10^{3}} \right ),
\end{eqnarray}
where $\sigma_{\rm cri}$ is the critical magnetization.

When the magnetar collapse, the emitting of spin-down wind stops. After the last spin-down wind passes the radius $r_{\rm d}$ (at time $t_{\rm b}$), steep decay begins.
As we proposed, the total X-ray light curve of the GRB with an internal plateau is composed of two parts:
(a) the brighter X-ray plateau and steep decay generated by to the dissipation of spin-down wind,
(b) the less luminous normal decay with decay index $\sim -1.2$ generated by the interaction between the GRB jet and its surrounding medium.
The steep decay should be powered by the residual magnetic energy in the spin-down wind.
Therefore, the dissipation of magnetic energy in the remnant spin-down wind follows the observed X-ray luminosity of the steep decay segment, i.e.,
\begin{eqnarray}\label{A2}
\frac{V}{4\pi}\frac{d(B^{2})}{dt}=L(t_{\rm b})\left ( \frac{t}{t_{\rm b}} \right )^{q},
\end{eqnarray}
 where $V=4\pi (ct)^{2}\Delta h$ is the volume of the spin-down wind with $t$ the time after the burst and $\Delta h$ the thickness of the spin-down wind,
$L(t_{\rm b})$ is the X-ray luminosity at time $t_{\rm b}$.
Note that $V\propto t^{2}$ and $\sigma\propto B^{2}$ (see equation \ref{eqs}), one has
\begin{eqnarray}\label{A3}
\frac{d\sigma }{dt}\propto t^{q-2}
\end{eqnarray}
from equation (\ref{A2}). According to equation (\ref{A3}), there is
\begin{eqnarray}\label{eq19}
\frac{\sigma_{\rm c}}{\sigma_{\rm d}}\sim \left (\frac{t_{\rm b}}{t_{\rm c}} \right )^{-q+1},
\end{eqnarray}
where $\sigma_{\rm c}$ is the magnetization of the spin-down wind at the collision, 
$t_{\rm c}$ is the time when the spin-down wind collides with the GRB jet.
Substituting equation (\ref{eq17}) into equation (\ref{eq19}) gives
\begin{eqnarray}\label{eq20}
\sigma_{\rm c}=\sigma_{\rm d} \left ( \frac{t_{\rm b}}{t_{\rm c}} \right )^{-q+1}
\end{eqnarray}
with
\begin{eqnarray}\label{eq21}
\sigma_{\rm d}=\left ( \frac{\eta(1+\sigma_{0})\Gamma _{\rm w}}{\Gamma_{\rm w,d}}-1 \right ).
\end{eqnarray}
Considering that the isotropic energy of some GRB X-ray plateaus can be as high as $\sim 10^{52}\rm erg$ (e.g., GRB 170714A, \citealt{2019ApJ...886...87D}),
$\eta$ at most $\sim 0.1$ since the most energy of the spin-down wind is released in X-ray emission.
On the basis of $-q>3$ for the steep decay, non-negligible mass loading (see equation \ref{eq18}) and equations (\ref{eq20}) and (\ref{eq21}),
to get $\sigma_{\rm c}>\sigma_{\rm cri}$, the value of $\sigma_{0}$ is too large to be reasonable.
Therefore, it is expected that the reverse shock always can be developed when the spin-down wind catches up with the GRB jet.
Correspondingly, a flare-like X-ray bump would be produced (see the lower panel of Figure \ref{fig1}, e.g., GRB 070110, \citealt{2007ApJ...665..599T}; and GRB 170714A, \citealt{2018ApJ...854..104H})
according to the internal shock model \citep{1994ApJ...430L..93R,1997ApJ...490...92K}.
Furthermore, since $\Gamma_{\rm w,d}\gg 1$ and $n_{\rm w}\ll n_{\rm jet}$ lead the reverse shock to be relativistic, we have $t_{\rm p}-t_{\rm c}\sim t_{\rm b}$ with
$t_{\rm p}$ the peak time of the X-ray bump (Figure \ref{fig1}).
So the internal energy release during the collision is $(1-\alpha)\eta L_{\rm sd}t_{\rm b}$ with $\alpha$
the value of $\Gamma_{\rm jet}/\Gamma_{\rm w,d}$ at $t_{\rm c}$,
i.e., the peak luminosity of the X-ray bump is $\sim (1-\alpha) \eta L_{\rm sd}$.

\begin{figure}[H]
 \centering
 \includegraphics[width=0.4\textwidth]{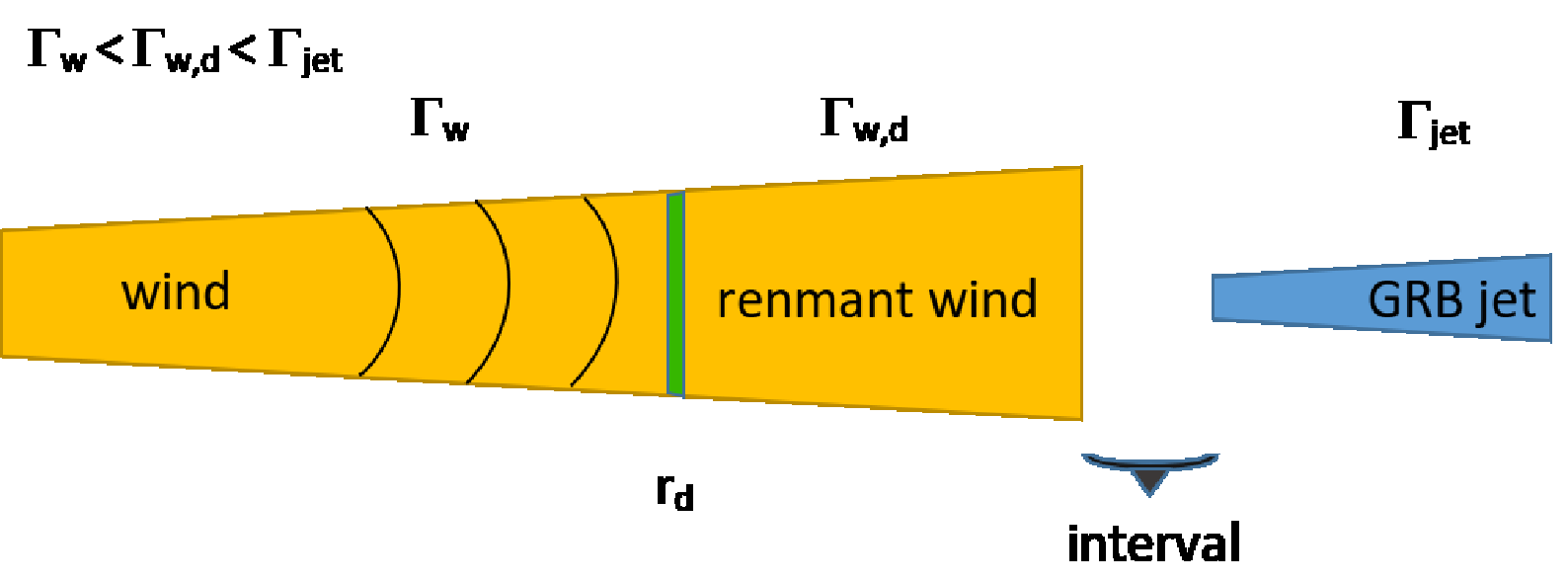}
 \includegraphics[width=0.4\textwidth]{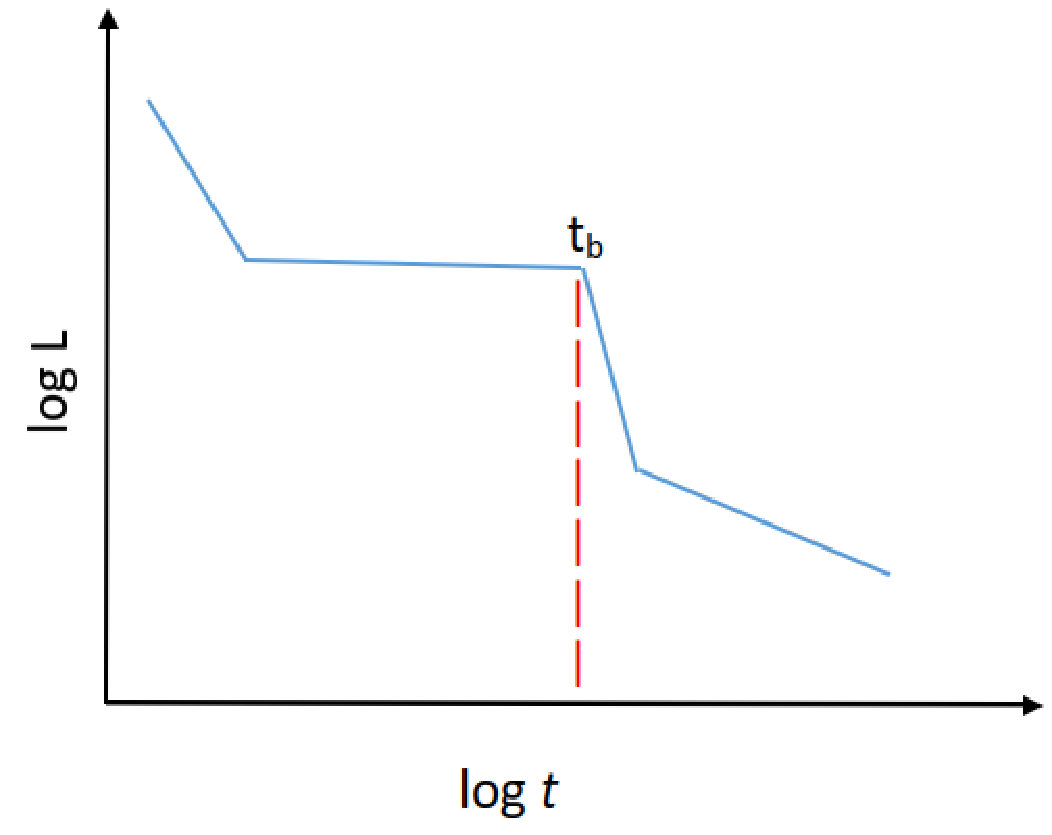}
 \caption{Schematic diagram of the relation between the spin-down wind and the GRB jet under $\Gamma_{\rm w}<\Gamma_{\rm w,d}<\Gamma_{\rm jet}$.
 Since there is $\Gamma_{\rm w,d}<\Gamma_{\rm jet}$, the remnant spin-down wind can not catch up with the GRB jet. The X-ray bump would not appear.}
  \label{geo}
\end{figure}

It's worth reminding that if the acceleration of the remnant spin-down wind is insufficient under the case discussed above,
i.e., $\Gamma_{\rm w}<\Gamma_{\rm w,d}<\Gamma_{\rm jet}$  (see the upper panel of Figure \ref{geo}),
the remnant spin-down wind would not catch up with the GRB jet. Therefore, no collision can appear, as well as the X-ray bump (see the lower panel of Figure \ref{geo}).

\begin{figure}[h]
 \centering
 \includegraphics[width=0.4\textwidth]{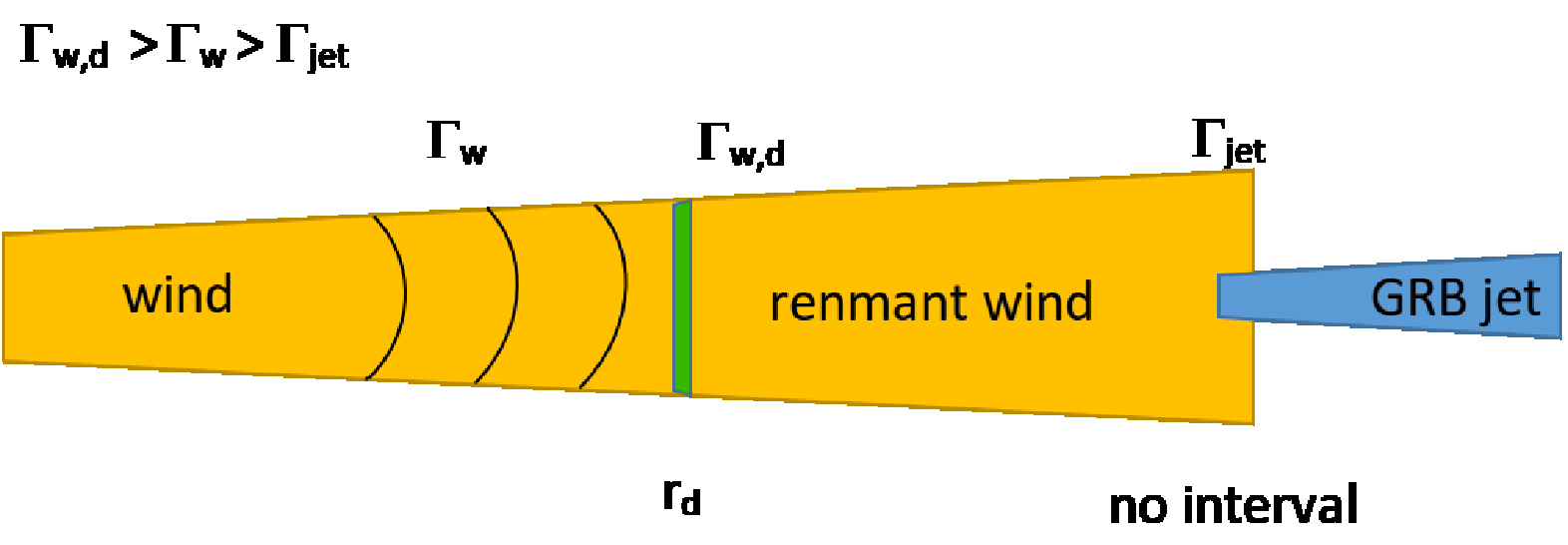}
 \includegraphics[width=0.4\textwidth]{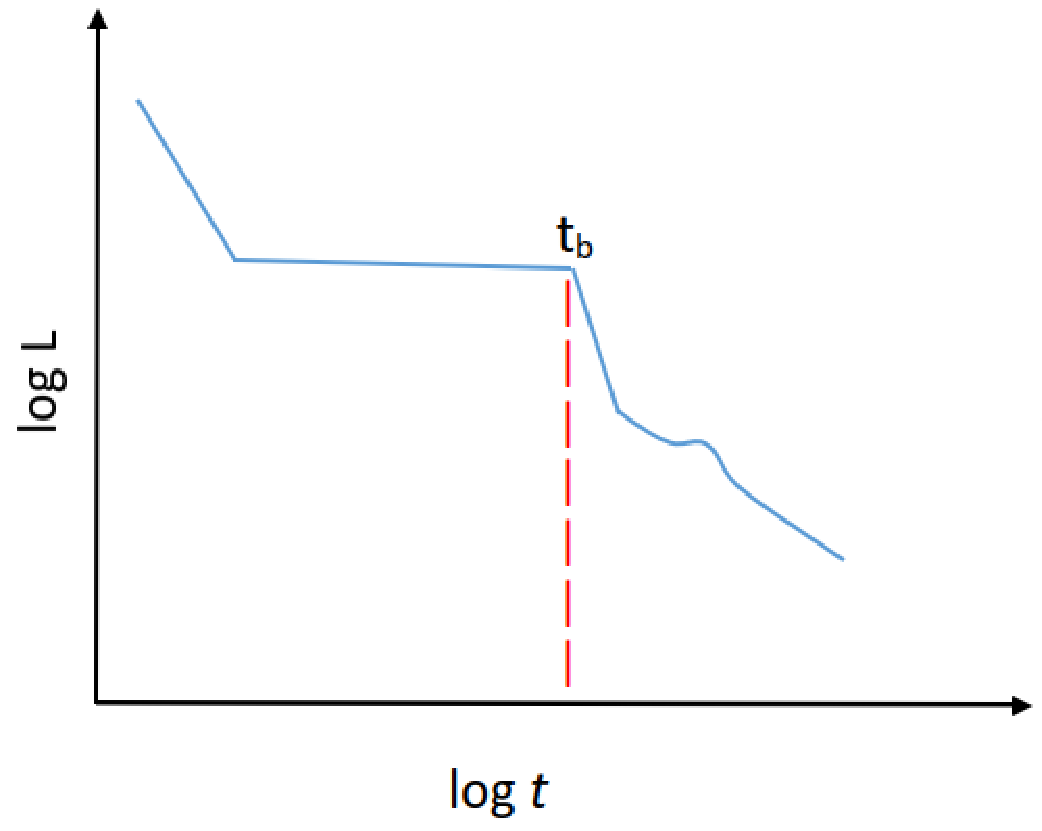}
 \caption{Schematic diagram of the relation between the spin-down wind and the GRB jet under $\Gamma_{\rm w,d}>\Gamma_{\rm w}>\Gamma_{\rm jet}$.
 The spin-down wind and GRB jet are connected together due to $\Gamma_{\rm w}>\Gamma_{\rm jet}$.
  When the spin-down wind reaches $r_{\rm d}$,
  the acceleration of the remnant spin-down wind is impeded by the GRB jet. So the distribution of the Lorentz factor of the remnant spin-down wind
  between the ``tail'' of the GRB jet and $r_{\rm d}$ is from $\Gamma_{\rm jet}$ to $\Gamma_{\rm w,d}$.
  This case is just the general energy injection discussed by  \cite{1998A&A...333L..87D} and \cite{2001ApJ...552L..35Z}.
  There will be a faint X-ray bump due to the powerless remnant spin-down wind (most of spin-down energy is released during the X-ray plateau segment).}
  \label{fig.3}
\end{figure}

On the other hand, if the spin-down wind and GRB jet are connected together until the spin-down wind reaches $r_{\rm d}$,
the acceleration of the spin-down wind can not be described by equation (\ref{eq17}).
Actually, when the GRB jet passes through $r_{\rm d}$ totally,
the corresponding distribution of the Lorentz factor of the remnant spin-down wind
between $r_{\rm d}$ and the ``tail'' of the GRB jet should be from $\Gamma_{\rm w,d}$ to $\Gamma_{\rm jet}$ (see the upper panel of Figure \ref{fig.3}).
Therefore, under this situation,
the interaction between the spin-down wind and GRB jet is more likely the general energy injection \citep{1998A&A...333L..87D,2001ApJ...552L..35Z}.
As the result of this interaction, only a faint X-ray bump can arise (see the lower panel of Figure \ref{fig.3}),
since the remnant energy of the spin-down wind, $\sim 10^{51}\;\rm erg$, is usually smaller than the kinetic energy of the GRB jet, $\sim 10^{52}-10^{54}\;\rm erg$.

\subsection{The independently evolved GRB jet and spin-down wind}
As discussed above, the dissipation of spin-down wind is independent of the GRB jet.
Since the opening angle of the spin-down wind (approximately isotropic) is much larger than that of the GRB jet ($\sim 10^{\circ}$),
there is a situation that only the radiation from the spin-down wind is observed.
In other words, an X-ray plateau without the corresponding GRB can be detected, such as, CDF-S XT2.
This could be an important phenomenon that can distinguish our model from the standard energy injection scenario,
since under the standard energy injection scenario, only when the spin-down wind injects into the GRB jet can the energy of the wind be converted into X-ray emission,
i.e., the X-ray plateau must be associated with a GRB.

On the other hand, GRBs are transient events, while spin-down winds from NSs will last for a long time.
These spin-down winds will exert long-lasting impacts on the evolutions of pulsar wind nebulae.
There is a conflict between theories and observations that the spin-down wind is usually dominated by the Poynting flux with $\sigma\gg 1$,
but the analyses of Crab nebula furnish $\sigma\ll 1$ (e.g., \citealt{1994ApJ...426..269B}, see \citealt{2016JPlas.82..0502P} for review).
The spin-down wind should be abruptly dissipated before reaching termination shock.
There is a problem that how does the strong magnetized spin-down wind become the weak magnetized wind.
As we suggested, part of the magnetic energy in the spin-down wind is converted into the kinetic energy of the spin-down wind before the wind reaching Crab nebula
(see also \citealt{1990ApJ...349..538C}).

These two different observations could be regarded as observational supports of our model.

\section{Discussion and Summary}\label{sec.3}
In the above, the acceleration of electrons during the magnetic energy release in the spin-down wind is not discussed,
since this is a complicated problem in astrophysics and beyond the scope of this paper.
Nonetheless, the acceleration of electrons through LAEMWs is not sensitive to $\sigma_{0}$, as long as $\sigma_{0}\gg 1$,
but depends on the ratio of spin-down power to the flux of electrons \citep{1994MNRAS.267.1035U}.
To get a suitable Lorentz factor (e.g., $\gamma_{\rm e}^{'}\sim 10^{2}$), the number density of electrons should satisfy equation (\ref{eq3}).

In this paper, we propose an improved proposal to explain the GRB internal plateaus under magnetar scenario
that the magnetized spin-down wind can almost be completely dissipated through LAEMWs behind the GRB jet.
Under this scenario, the interaction between the remnant spin-down wind
and the jet may result in a flare-like or a faint X-ray bump following the steep decay.
Especially, the acceleration mechanism of electrons in the spin-down wind under our scenario can be
different with that of the standard external shock scenario,
so that the multi-band afterglow light curve of these two may show different features.
For example, during the plateau segment (including ordinary plateau, if the magnetar is stable),
the multi-band afterglow of the former is chromatic, however the latter is more likely achromatic.
This difference may be useful for distinguishing the origins of the plateaus.

It is worth noting that although we claim that the spin-down wind can be dissipated via LAEMWs,
the proposal that the spin-down power is not injected into the forward external shock
but continuously dissipated behind the GRB jet is compatible with other mechanisms as long as
they can dissipate the spin-down wind instantaneously at a certain distance from the GRB central engine.
For example, if the Lorentz factor has an appropriate distribution in the spin-down wind (e.g., with a slower head and faster tail due to time-evolution mass loading),
the spin-down wind will shrink enough during travelling, then there may be a certain time that
the striped magnetic field can be dissipated through the magnetic reconnection induced by the self-compression of the wind.
We just believe that LAEMWs are a much simpler and more natural mechanism to power the X-ray plateaus in GRB physics.

Determining the origin of the plateaus has more than just astronomical implications.
It is also important for understanding the equation of state of supranuclear matter,
a problem relevant to non-perturbative quantum chromo-dynamics (QCD, see, e.g., \citealt{2018Sci...61....531X}).
Although we have entered the era of multi-messenger astronomy, the current gravitational-wave detectors
can not provide effective information of the remnant of GW170817-like event.
However, GRB X-ray plateaus may provide opportunities to achieve this goal, as well as
the cognition of the non-perturbative QCD, since the observational behavior of plateau is strongly
related to the life of a supramassive/hypermassive NS and thus to the stiffness of the equation of state (i.e., $M_{\rm TOV}$).

\section{Acknowledgement}
We would like to thank the anonymous referee for his/her very
useful comments that have allowed us to improve our paper.
We would like to thank Profs. Yongquan Xue and Huirong Yan for useful discussions and valuable comments.
We would like to thank Dr. Xuhao Wu for helping me with the English expression.
We thank Xianggao Wang for helping us to find the sample, GRB 150910A, to test our model after we explain our model to him.
This work was supported by the National Key R\&D Program of China (Grant
No. $\rm 2017YFA0402602$), the National Natural Science Foundation of China
(Grant Nos. $11673002$, and $\rm U1531243$), and the Strategic Priority Research
Program of Chinese Academy Sciences (Grant No. $\rm XDB23010200$).
\\
\\

\end{document}